\documentclass[showkeys,twocolumn,aps]{revtex4}
\usepackage{amsmath,amssymb}
\usepackage{graphicx}% Include figure files

\begin{document}

%
% \preprint{\begin{minipage}{2in}\begin{flushright}
%   SLAC-PUB-7121 (T) \\[-3mm] hep-lat/9603016 \end{flushright}
%   \end{minipage}}
%\input epsf
% \draft
% \twocolumn[\hsize\textwidth\columnwidth\hsize\csname@twocolumnfalse\endcsname
%

%\title{Soluciones de las ecuaciones de apareamiento de Richardson en representaciones con energ\'ias complejas}
\title{Comparaci\'on entre las soluciones de Lipkin-Nogami y Richardson con energ\'ia compleja en el modelo de Lipkin \\
		 \vspace{2mm}
		 Comparison between the Lipkin-Nogami and Richardson solutions with complex single particle energy in the Lipkin's model}

\author{Rodolfo M. Id Betan\footnote{e-mail: idbetan@ifir-conicet.gov.ar}}

\affiliation{Instituto de F\'{\i}sica Rosario (CONICET), Ocampo y Esmeralda, S2000EZP Rosario. \\ 
             Departamento de F\'{\i}sica. FCEIA (UNR), Av. Pellegrini 250, S2000BTP Rosario.\\
%             Departamento de F\'{\i}sica y Qu\'{\i}mica. Facultad de Ciencias Exactas, Ingenier\'{\i}a y Agrimensura (UNR), Av. Pellegrini 250, S2000BTP Rosario.\\
             Instituto de Estudios Nucleares y Radiaciones Ionizantes (UNR), Riobamba y Berutti, S2000EKA Rosario.}
%             Instituto de Estudios Nucleares y Radiaciones Ionizantes (UNR), Riobamba y Berutti, S2000EKA Rosario. Argentina.}
%
%------------------------------------
%comienza abtract en español e inglés             
\begin{abstract}
 \begin{center}
   \renewcommand{\date}{}
   \date{Recibido: xx/1/17; aceptado: xx/xx/17}
 \end{center}
La interacci\'on de pairing es de gran inter\'es debido a que es una de las componentes principales de la interacci\'on residual en sistemas de muchos cuerpos. Las aproximaciones de Bardeen-Cooper-Schieffer (BCS) y Lipkin-Nogami (LN) dan soluciones aproximadas al Hamiltoniano de pairing. Por otro lado, el pairing constante, admite soluci\'on exacta, llamada soluci\'on de Richardson. Los n\'ucleos alejados de la l\'inea de estabilidad tienen importantes correlaciones con los estados de dispersi\'on y por ende, los estados del continuo deben ser tenidos en cuenta en forma expl\'icita en la descripci\'on de tales sistemas. Una forma de incluir el continuo es a trav\'es de los estados de Gamow, este es, soluciones de la ecuaci\'on de Schroedinger con energ\'ia compleja. En este trabajo comparamos las soluciones aproximadas de BCS y LN con la soluci\'on exacta en el modelo sim\'etrico de Lipkin con energ\'ia compleja. Encontramos que la soluci\'on de LN resulta muy parecida a la exacta y que la extensi\'on de la soluci\'on de la BCS al plano de energ\'ia complejo produce soluciones puramente imaginarias para el gap cuando $G<G_c$.
 \vskip 0.3cm
 \noindent
 {\footnotesize
    \renewcommand{\keywords}{Palabras claves: }
 }%end footnotesize
 \vskip 0.5cm
 \noindent
The pairing interaction is one of the most important contribution of the residual interaction and then, it is of major importance for the study of many-body systems. One can get solutions of the pairing Hamiltonian throught the Bardeen-Cooper-Schieffer (BCS) or the Lipkin-Nogami (LN) approximations but, the pairing Hamiltonian admit exact solution worked out by Richardson. Nuclei far away from the stability line have important correlations with the continuum part of the energy spectrum, due that the Fermi level is very close to the contiuum thershold. The correlations with the continuum can be included in the many-body description through the complex energy states, called Gamow states. In this work we compare the approximates and exact solutions of the pairing Hamiltonian in real and complex-energy representations. In the application of this formulation to the symmetric Lipkin model, we found that the LN solution is in a good agreement with the exact one; besides, the extension of the BCS solution to the complex energy plane gives solution even for strength below the critical one, which is purely imaginary.
\end{abstract}
\maketitle
%\vskip 0.5cm]
% 
%%%%%%%%%%%%%%%%%%%%%%%%%%
%\section{INTRODUCCI\'ON}
\section{Introduction}
Many-body configurations in the continuum is an important issue to understand the properties of loosely bound systems, as for example, nuclei close to the drip line \cite{2003Okolowicz}. The complex energy poles of the scattering matrix correspond to complex energy eigenvalue of a single particle Hamiltonian with purely outgoind boundary condition. They are called Gamow states and they represent decay states in the continuum \cite{2009Michel}. They have information of the structure of the real energy continuum spectrum: the real part of the complex pole gives the resonant energy while the reciprocal of its imaginary part is proportional to half-live of the unstable state \cite{1988Kukulin}.

The constant pairing interaction, although simple, is an important component of the particle-particle interaction \cite{1964Lane,2003Dean}.  The pairing Hamiltonian can be solved using the Bardeen-Cooper-Schrieffer (BCS) approximation \cite{1957Bardeen}; but, for small number of particles, this approximation is not a satisfactory solution. An improve solution was given by Nogami \cite{1964Nogami,1965Nogami} using the technique developed by Lipkin ir Ref. \cite{1960Lipkin}, now known as Lipkin-Nogami (LN) approximation. The solutions in the LN approximation including the continnum single particle density was work out recently in Ref. \cite{2017IdBetan}.  But, the constant pairing has exact solution worked out by Richardson \cite{1963Richardson,1964Richardson}. The eigenfunctions of the BCS and LN  solutions do not conserve the number of particles of the system, while the Richardson solution does.

A test system, which is non trivial but it is simple enough to be exactly solvable was given by Lipkin, Meshkov and Glick \cite{1965Lipkin}, now know as Lipkin model or symmetric model. It is used to test the validity of new formalisms and techniques as well as to illustrate more complicated models in many-body systems \cite{1961Hogaasen,1965Mechkov,1973Pradhan,2004Ring}.

In this work we calculated the many-body energy in the BCS, LN and Richardson frameworks with complex energy in the Lipkin model. The Lipkin formalism is described  in section \ref{sec.lf}. In section \ref{sec.formalism} the BCS, LN and Richardson solutions are given. The application to the Lipkin model is presented in section \ref{sec.applications}. Finally, in the last section \ref{sec.conclusions} we draw some conclusions.

%%%%%%%%%%%%%%%%%%%%%%%%%%%%%%%%%%%%%%%%%%
%\section{FORMALISMO}\label{sec.formalismo}
%\section{Formalism}\label{sec.formalismo}
\section{Lipkin formalism}\label{sec.lf}
Let us assume that $| \Phi \rangle$ is an approximate solution of the unknown exact ground state of a many-body system which is described by a Hamiltonian $H$. The wave function (w.f.) $| \Phi \rangle$ describes nicely some properties of the system but, at the same time, it violates some other property $\hat{S}$. For example, the BCS w.f. $| \Phi_{BCS} \rangle=\prod_{k>0}(1+\phi_k)a_k^\dagger a_{\bar{k}}^\dagger |0 \rangle$ \cite{1957Bardeen} describes nicely the pairing property of the many-body system but it is not an eigenfunction of the particle number operator $\hat{N}$. We may expand $|\Phi \rangle$ in a basis $|\phi_S \rangle$ of eigenfunctions of $\hat{S}$, i.e., $| \Phi \rangle = \sum_S c_S | \phi_S \rangle$, with $\langle \Phi | \Phi \rangle=1$ ($\sum_S c_S^2=1$ and $c_S$ real numbers.) In our example $| \Phi_{BCS} \rangle =\sum_{N=0}^{N_{max}} c_N | \phi_N \rangle$ were $N$ is even, $N_{max}$ is the maximum particle number allowed by the representation and $\hat{N} | \phi_N \rangle=N | \phi_N \rangle$. 

An approximation of the ground state energy would be $\langle \Phi | H | \Phi \rangle$, but since $| \Phi \rangle$ does not conserve the property $\hat{S}$, we hope that $\langle \phi_S | H | \phi_S \rangle$ will be a better approximation for some specific value of $S$ of the observable $\hat{S}$. The trick consist to use the approximate w.f. $| \Phi \rangle$ (called model w.f.), which we assume it is easier to handle, to obtain $\langle \phi_S | H | \phi_S \rangle$, together with a model Hamiltonian $\mathcal{H}$ (to be build) for which our model w.f. is an eigenvector  \citep{1960Lipkin}.

We defined our model Hamiltonian as
\begin{equation}
  \mathcal{H}=H-f(\hat{S}) \, ,
\end{equation}
with $f(\hat{S}) | \phi_S \rangle = f(S) | \phi_S \rangle$, then
\begin{equation}\label{eq.ph1}
     \langle \Phi | \mathcal{H} | \Phi \rangle =
           \langle \phi_S | H - f(S) | \phi_S \rangle
\end{equation}

If $f(\hat{S})$ in is chosen in such a way that $|\phi_S \rangle$ are all degenerate eigenfunctions of $\mathcal{H}$, i.e. $\mathcal{H}|\phi_S \rangle=\mathcal{E}|\phi_S \rangle$ then 
\begin{equation}\label{eq.ph2}
     \langle \Phi | \mathcal{H} | \Phi \rangle = \mathcal{E} \, .
\end{equation}

By combining Eqs. (\ref{eq.ph1}) and (\ref{eq.ph2}) we get,
\begin{equation}
   \mathcal{E} = \langle \phi_S | H - f(S) | \phi_S \rangle\, ,
\end{equation}
notice that $\mathcal{E}$ is not the eigenvalue of our system but,
\begin{equation}
  \langle \phi_S | H | \phi_S \rangle = \mathcal{E} + f(S)\, ,
\end{equation}
then, the above discussion assume that $\mathcal{E}=\langle \Phi | \mathcal{H} | \Phi \rangle$ is easier to calculate than $\langle \phi_S | H | \phi_S \rangle$, i.e., it is easier to solve the eigenvalue problem for the model Hamiltonian $\mathcal{H}|\Phi \rangle$ than the original one $H|\phi_S \rangle$. 

Probably, the only exactly known $f(\hat{S})$ is the momentum operator, in all the other cases this function has to be approximated by a Taylor's series, with the hope that a few terms will be enough to reproduce the truly ground state energy. So, let us assume that 
\begin{equation}\label{eq.taylor}
   f(\hat{S})=f_1 \hat{S}+f_2 \hat{S}^2+\dots \, .
\end{equation}
In the case that one truncates the series, the condition that $\langle \phi_S | \mathcal{H} | \phi_S \rangle$ be degenerated for all $S$ is not fulfill. In such a case one must complement the problem with some other subsidiary condition.

If $\hat{S}$ represents the particle number operator $\hat{N}$, the simplest approximation of $f(\hat{N})$ is when we keep the first term of the series, $f(\hat{N})=\lambda \hat{N}$, then $\mathcal{H}=H-\lambda \hat{N}$, i.e. the BCS approximation. Then $\langle \Phi_{BCS} | \mathcal{H} | \Phi_{BCS} \rangle = \mathcal{E}_{BCS}$ and the subsidiary condition is that the mean value of the particle number operator is fixed, hence $\langle \phi_N | H |\phi_N \rangle = \mathcal{E}_{BCS} + \lambda N$. The condition $\langle \Phi_{BCS} | \hat{N} | \Phi_{BCS} \rangle = N$ determines the value of the parameter $\lambda$.

The equation 
\begin{equation}
   \langle \phi_S | H | \phi_S \rangle = \mathcal{E} + f(S)
    = \langle \Phi | \mathcal{H} | \Phi \rangle + f(S)
\end{equation}
can be rearranged to be written,
\begin{eqnarray}
   \langle \phi_S | H | \phi_S \rangle &=& \langle \Phi | H | \Phi \rangle 
           + f_1 (S - \langle \Phi | \hat{S} | \Phi \rangle) \nonumber \\
           && + f_2 (S^2 - \langle \Phi | \hat{S}^2 | \Phi \rangle) + \dots
\end{eqnarray}
where each term can be interpreted as a correction term. This way of writing the mean value $\langle \phi_S | H | \phi_S \rangle$ is a bit tricky. For example, in the BCS example
\begin{eqnarray}
  \langle \phi_N | H | \phi_N \rangle &=& \langle \Phi_{BCS} | H | \Phi_{BCS} \rangle \nonumber \\
       && + \lambda ( N - \langle \Phi_{BCS} | \hat{N} | \Phi_{BCS} \rangle)
\end{eqnarray}
and due the subsidiary condition $\langle \Phi_{BCS} | \hat{N} | \Phi_{BCS} \rangle=N$ the correction would be zero. The point we must remember is that we built the model Hamiltonian because is was easy to manipulate with our model wave function, i.e. we don't solve the eigenvalue problem $\langle \Phi | H | \Phi \rangle$, instead we solve the eigenvalue problem $\langle \Phi | \mathcal{H} | \Phi \rangle$. In this way we never face terms of the form $f_i(S^i-\langle \Phi | \hat{S^i} | \Phi \rangle)$ which could be zero. In our BCS example it means that we don't solve $\langle \Phi_{BCS} | H | \Phi_{BCS} \rangle$ but $\langle \Phi_{BCS} |H - \lambda \hat{N} | \Phi_{BCS} \rangle$. 

The next step is to find a systematic way to obtain the parameters $f_i$ for $i\ge1$ which does not involve the states $|\phi_S \rangle$ but instead involves the model w.f. $| \Phi \rangle$. The model w.f. and the model Hamiltonian satisfies the relation 
\begin{equation}\label{eq.condition}
  \langle \Phi | \mathcal{H} \, g(\hat{S}) | \Phi \rangle
        = \langle \Phi | \mathcal{H} | \Phi \rangle \,  
           \langle \Phi | g(\hat{S}) | \Phi \rangle
\end{equation}
for any function $g(\hat{S})$. We can choose a set of functions $g_i(\hat{S})=\hat{S}^i$ with $i=1,2,\dots$ in order to evaluate the coefficients $f_i$. Then, a self-consistency conditions (independent of $|\phi_S \rangle$) is obtained by rearranged Eq. (\ref{eq.condition}) with $g(\hat{S})$ replaced by $\hat{S}^i$,
\begin{equation}
   \langle \Phi | \mathcal{H} \, ( \hat{S}^i - \langle \Phi | \hat{S}^i | \Phi \rangle) |\Phi \rangle = 0
\end{equation}

The application of this condition to our BCS example would give, 
\begin{equation}
   \langle \Phi_{BCS} | \mathcal{H} \, 
   				( \hat{N} - \langle \Phi_{BCS} | \hat{N} | \Phi_{BCS} \rangle) | \Phi_{BCS} \rangle=0
\end{equation}
and then
\begin{equation}
   \langle \Phi_{BCS} | \mathcal{H} \, \hat{N} | \Phi_{BCS} \rangle
         - \mathcal{E}_{BCS} \langle \Phi_{BCS} | \hat{N} | \Phi_{BCS} \rangle =0 
\end{equation}
By inserting $| \Phi_{BCS} \rangle \langle \Phi_{BCS}|$ between $\mathcal{H} \, \hat{N}$ and using the condition $\mathcal{H}_{20}=0$ \cite{1971Fetter,1966Goodfellow} we get
\begin{equation}
    \mathcal{E}_{BCS} \langle \Phi_{BCS} | \hat{N} | \Phi_{BCS} \rangle
        -  \mathcal{E}_{BCS} N = 0
\end{equation} 
which gives the standard condition used in BCS, $\langle \Phi_{BCS} | \hat{N} | \Phi_{BCS} \rangle = N$.

%%%%%%%%%%%%%%
\section{Model solutions}\label{sec.formalism}
The constant pairing Hamiltonian reads,
\begin{eqnarray}\label{eq.h}
   H &=& H_{sp} + V \, ,
\end{eqnarray}
where
\begin{eqnarray}
    H_{sp} &=& \sum_j \epsilon_j \hat{n}_j \hspace{5mm} \hat{n}_j= \sum_m a^\dagger_{jm} a_{jm}   \nonumber \\
    V &=&-g\; P^\dagger P \hspace{5mm} P^\dagger
          = \sum_{j m>0} a^\dagger_{jm} a^\dagger_{j \bar{m}} 
\end{eqnarray}
with $a^\dagger_{j \bar{m}} \equiv (-)^{j-m} a^\dagger_{j, -m}$, and $g$ the strength of the interaction. The particle number operator is $\hat{N}=\sum_j \hat{n}_j$

%%%%%%%%%%%%%%%%%%%%%%%
\subsection{Non conserving particle number solutions}
In this section we will applied the Lipkin method \cite{1960Lipkin} of section \ref{sec.lf} to obtain approximate solutions of the pairing Hamiltonian (\ref{eq.h}).

\subsubsection{BCS solution}
The Taylor's expansion Eq. (\ref{eq.taylor}) in the particle number operator $\hat{N}$ up to first order defines the usual BCS model Hamiltonian
\begin{equation}
   \mathcal{H}_{BCS} = H - \lambda \hat{N} \, ,
\end{equation}
while the model w.f. is defined as \cite{2007Suhonen}
\begin{equation} \label{eq.wfbcs}
    | \Phi_{BCS} \rangle = \prod_{m>0}
             \left( u_j + v_j  a_j^\dagger a_{\bar{j}}^\dagger \right) |0 \rangle
\end{equation}
with the coefficients $u_j$ and $ v_j$ satisfying $u_j^2 + v_j^2=1$. The ground state energy in this approximation is
\begin{eqnarray}\label{eq.bcs}
     E_{BCS} &=& \langle \Phi_{BCS} | \mathcal{H}_{BCS} | \Phi_{BCS} \rangle + \lambda N \\
                      &=& \sum_{jm} (\epsilon_j - \frac{g}{2} v_j^2)v_j^2 - \frac{\Delta^2}{g}
\end{eqnarray}  
with 
\begin{eqnarray}
  v_j^2 &=& \frac{1}{2} \left( 1 - \frac{e_j}{E_j} \right) \nonumber \\
  E_j &=&  \sqrt{e_j^2 + \Delta^2} \nonumber \\
  e_j &=& \epsilon_j - \lambda - g \; v_j^2 
\end{eqnarray}

The gap parameter $\Delta$ and the Fermi level $\lambda$ are obtained by solving the following system of equations
\begin{eqnarray}
 \frac{4}{g} &=& \sum_j \frac{(2j+1)}{E_j} \\
  N&=&\sum_j (2j+1) v^2_j 
\end{eqnarray}

\subsubsection{Lipkin-Nogami solution}
The model Hamiltonian we obtain by taking the Taylor's expansion Eq. (\ref{eq.taylor}) in the particle number operator $\hat{N}$ up to the second order defines the Lipkin-Nogami (LN) model Hamiltonian \cite{1964Nogami}
\begin{equation}
   \mathcal{H}_{LN} = H - \lambda_1 \hat{N} - \lambda_2 \hat{N}^2 \, ,
\end{equation}
with $H$ as Eq. (\ref{eq.h}). The model w.f. $| \Psi_{LN} \rangle$ is like Eq. (\ref{eq.wfbcs}) but with different coefficients $u_j$ and $v_j$. They are determined in terms of the parameters $\Delta$, $\lambda_1$ and $\lambda_2$ by solving the following system of three equations
\begin{eqnarray}
 \frac{4}{g} &=& \sum_j \frac{(2j+1)}{E_j}  \\
  N&=&\sum_j (2j+1) v^2_j  \\
 \frac{4\lambda_2}{g} &=& \frac{(\sum_{jm} u_j^3 v_j)(\sum_{jm} u_j v_j^3)- 2 \sum_{jm} (u_j v_j)^4}
                            {(\sum_{jm} (u_j v_j)^2)^2 - 2\sum_{jm} (u_j v_j)^4} \nonumber \\
\end{eqnarray}
with $\sum_{m}= 2j+1$ and
\begin{eqnarray}
  v_j^2 &=& \frac{1}{2} \left( 1 - \frac{e_j}{E_j} \right) \nonumber \\
  E_j &=&  \sqrt{e_j^2 + \Delta^2} \nonumber \\
  e_j &=& \epsilon_j - \lambda + (4 \lambda_2 - g) v_j^2 \nonumber \\
  \lambda &=& \lambda_1 + 2 \lambda_2 (N+1) \nonumber
\end{eqnarray}

The ground state energy is
\begin{eqnarray}\label{eq.ln}
     E_{LN} &=&  \langle \Phi_{LN} |\mathcal{H}_{LN} |\Phi_{LN} \rangle
          +  \lambda_1 N + \lambda_2 N^2 \nonumber \\
                &=& \sum_{jm} (\epsilon_j - \frac{g}{2} v_j^2)v_j^2 - \frac{\Delta^2}{g} - \lambda_2 \sum_{jm} 2 u_j^2 v_j^2 %\nonumber \\
\end{eqnarray}

%%%%%%%%%%%%%%%%%%%%%
\subsection{Conserving particle number solution}
The conserving particle number solution for a system of even $N$ fermions is given in terms $N_{pairs}=N/2$ parameter $E_n$ called pair energies. These parameters are obtained by solving a system of $N_{pairs}$ equations, called Richardson's equations \citep{1963Richardson,1964Richardson}
\begin{equation}
    \frac{1}{g} - \frac{1}{2} \sum_j \frac{2j+1}{2\epsilon_j - E_n} 
               + 2 \sum_{n'=1 \\, n' \ne n}^{N_{pairs}} \frac{1}{E_{n'} - E_n}=0
\end{equation}

The many-body ground state energy $E_{Rich}$ is defined taken the lowest $N_{pairs}$ pair-energy $E_n$
\begin{equation}
 E_{Rich} = \sum_n^{N_{pairs}} E_n
\end{equation}

%%%%%%%%%%%%%%%%%%%%%%%%%%%%%%%%%%%%%%%%%%%%
%\section{APLICACI\'ON}\label{sec.aplicaciones}
%aplicaci\'on
%\begin{figure}[ht]
%\begin{center}
%%\includegraphics[width=0.45\textwidth]{ge_lmax4}
%\caption{caption ...}
%\label{fig.ge}
%\end{center}
%\end{figure}
\section{Application: Lipkin model}\label{sec.applications}
The Lipkin model, also called symmetric model, consist of two equally degenerate levels with energies $\epsilon_u$ and $\epsilon_d$ at half filling. The notation of the previous section reduces to $\sum_{jm}=\sum_{u,d} \sum_{m=-\Omega}^{\Omega}$. Hence, the degeneracy for each level is $2\Omega$ and $N=2\Omega$. Let us introduce the parameter $\epsilon>0$
\begin{equation}
    \epsilon=\epsilon_u-\epsilon_d \, ,
\end{equation} 
which defines the energy separation between the two levels.  The following values are used for the applications:
\begin{eqnarray}
   N &=& 10 \nonumber \\
   \epsilon_d &=& -0.5\, MeV \nonumber \\
   \epsilon_u &=& (0.5-i\, \gamma) \, MeV \nonumber 
\end{eqnarray}

By using the relations of the previous section we found the following algebraic solutions for the BCS and LN approximations:

\paragraph{BCS solution:}
\begin{eqnarray}
   \lambda &=& \frac{\epsilon_u+\epsilon_d}{2} - \frac{g}{2} \nonumber \\
   E &=& E_u = E_d= g \; \Omega \nonumber \\
   \Delta &=&  g \; \Omega \sqrt{1 - \frac{\epsilon^2}{(2g\Omega-g)^2}} \nonumber
\end{eqnarray}
with 
\begin{equation}\label{eq.gc1}
   g > g_c \equiv \frac{\epsilon}{2\Omega-1}
\end{equation}

The ground state energy $E_{BCS}$ Eq. (\ref{eq.bcs}), relative to the non-interacting system is
\begin{equation}
 E_{BCS} - 2\Omega \epsilon_d = \Omega \left( \epsilon - \frac{g}{2} \right) 
     - \frac{\Delta^2}{g} 
    -  \frac{\Omega e}{E} \left( \epsilon + \frac{g}{2}\, \chi \right)
\end{equation}
where
\begin{equation}
   \chi=\frac{\epsilon}{2\Omega g - g}
\end{equation}

\paragraph{LN solution:} The Lipkin-Nogami solution can also be obtained analytically,
\begin{eqnarray}
   \lambda &=& \frac{\epsilon_u+\epsilon_d}{2} + \frac{\alpha}{2}; 
         \hspace{5mm}
         \alpha=4 \lambda_2 - g >0 \nonumber \\
    E &=& E_u=E_d= g \; \Omega \nonumber \\
     \Delta &=&  g \; \Omega \sqrt{1 - \frac{\epsilon^2}{(2g\Omega+\alpha)^2}} \nonumber
\end{eqnarray}
with
\begin{equation} \label{eq.gc2}
   g > g_c \equiv \frac{\epsilon-\alpha}{2\Omega}
\end{equation}

Since $4 \lambda_2 - g >0$ we are interested in the positive $\alpha$ solution of the following cubic equation (see also Eq. (15) in Ref. \cite{1965Nogami})
\begin{equation} \label{eq.alpha}
 \alpha(2\Omega-1)[(2g\Omega+\alpha)^2 - \epsilon^2] - 2g\Omega \epsilon^2 = 0
\end{equation}

\begin{figure}[htb]
\vspace{5mm}
   \includegraphics[angle=-90,width=0.45\textwidth]{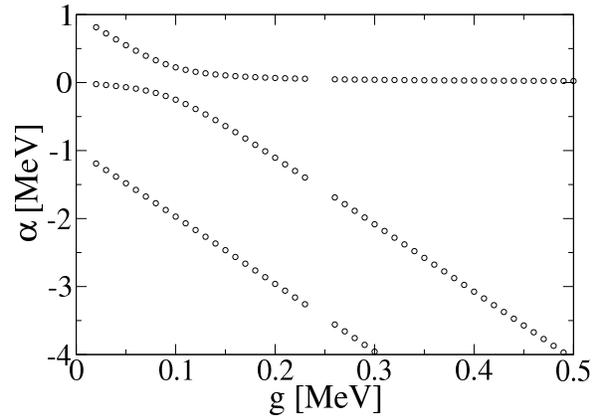}
   \caption{\label{fig.alpha} Solutions of the cubic equation (\ref{eq.alpha}) $g \in [0,0.5]$ MeV. Missing points in the figure is because no convergence was found.}
\vspace{5mm}
\end{figure}

Figure \ref{fig.alpha} shows the three possible solutions of Eq. (\ref{eq.alpha}) for each value of the strength $g$ in the range $[0,0.5]$ MeV. As the strength goes to zero, the parameter $\alpha$ goes to zero and to $\pm \epsilon$ (with $\epsilon=1$ MeV). 

The ground state energy $E_{LN}$ Eq. (\ref{eq.ln}) relative to the non-interacting Fermi energy $2\Omega \epsilon_d$ gives,
\begin{eqnarray}
    E_{LN} - 2\Omega \epsilon_d &=&
            \Omega \left( \epsilon - g - \frac{\alpha}{2} \right) - \frac{\Delta^2}{g} 
    -  \Omega \chi \left( \epsilon - \frac{\alpha}{2}\, \chi \right) \nonumber \\
\end{eqnarray}
where
\begin{equation}
 \chi=\frac{\epsilon}{2\Omega g + \alpha}
\end{equation}

\begin{figure}[htb]
\vspace{10mm}
   \includegraphics[angle=-90,width=0.45\textwidth]{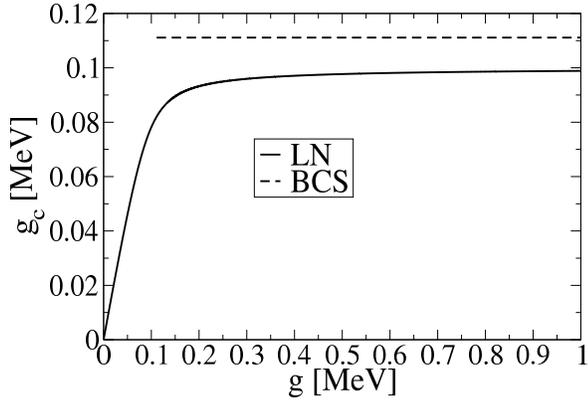}
   \caption{\label{fig.gc} $g_c$ of Eqs. (\ref{eq.gc1}) and (\ref{eq.gc2}) (for $\alpha >0$) versus g for the BCS and LN approximations, respectively.}
\vspace{5mm}
\end{figure}

Equations (\ref{eq.gc1}) and (\ref{eq.gc2}) show that in order to find solution for the BCS and LN approximations, respectively, the strength $g$ has to be greater than $g_c$. Figure \ref{fig.gc} shows the value of $g_c$ as a function of $g$ for $\alpha >0$. It is found that in order to have non trivial solution in the BCS approximation $g$ has to be bigger that a threshold value, while the LN approximation has no trivial solution for any value of the strength.

\begin{figure}[ht]
\vspace{5mm}
   \includegraphics[angle=-90,width=0.45\textwidth]{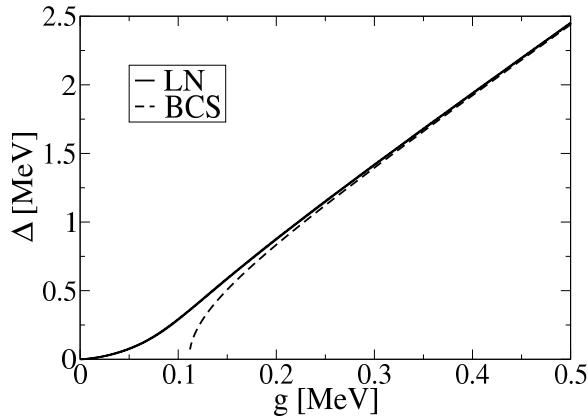}
   \caption{\label{fig.gap} Pairing gap versus $g$.}
\vspace{5mm}
\end{figure}

Figure \ref{fig.gap} extends the comparison done in Fig. 1 of Ref. \cite{1965Nogami} between the BCS and LN gap parameter $\Delta$ as a function of the strength $g$ for $\alpha >0$ and $\gamma=0$ to stronger strength. They compare well for strong correlation but they depart each other for small value of the strength. The figure also shows the nonphysical behavior of the pairing gap in the BCS approximation, i.e. $\Delta=0$ for $g\lesssim 0.1$ MeV.

\begin{figure}[ht]
\vspace{5mm}
    \includegraphics[angle=-90,width=0.45\textwidth]{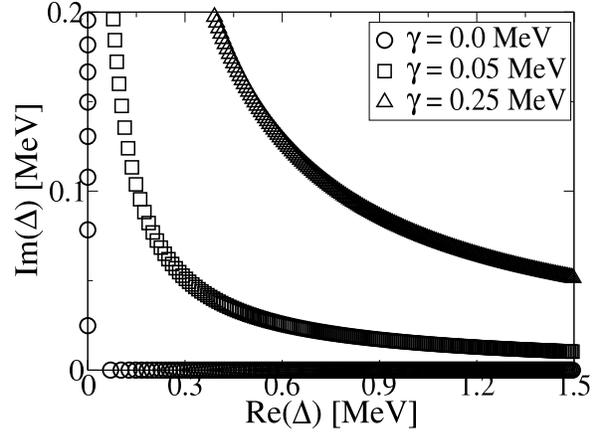}
    \caption{\label{fig.gapcx} Real and imaginary part of the pairing parameter calculated in the BCS approximation for different value of $\gamma$ for $g<0.32MeV$.}
\vspace{5mm}
\end{figure}

In Fig. \ref{fig.gapcx} we show the gap parameter as complex magnitude in the BCS approximation for $\gamma=0$, $0.05$, and $0.25$ MeV. For complex energy the gap parameter is also complex but for $\gamma=0$ the gap is purely real up to a minimum value and then it becomes purely imaginary, i.e., the trivial solution appears here as a complex solution with $\Delta$ purely imaginary; hence, in the complex plane the constant gap has no trivial solution for any value of $g$.

\begin{figure}[htb]
\vspace{5mm}
   \includegraphics[angle=-90,width=0.45\textwidth]{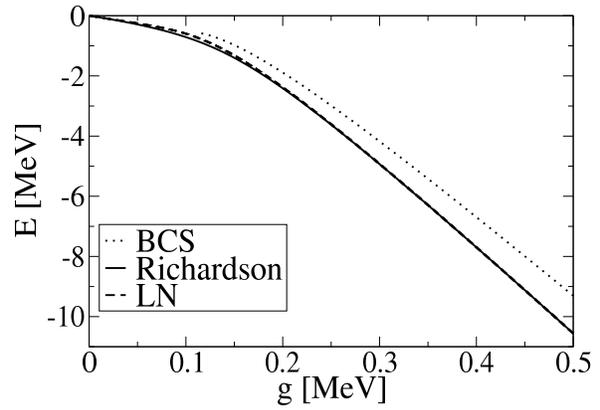}
   \caption{\label{fig.ereal} Binding energy relative to the non-interacting Fermi sea versus $g$ for $\gamma=0$.}
\vspace{5mm}
\end{figure}

The ground state energy relative to the free system of the non-conserving particles solutions are compared with the exact Richardson solution in Fig. \ref{fig.ereal} for $\gamma=0$. We observe a very good agreement between the approximate LN and the exact (Richardson) solutions  for all values of the pairing strength.

\begin{figure}
\vspace{2mm}
    \includegraphics[angle=-90,width=0.45\textwidth]{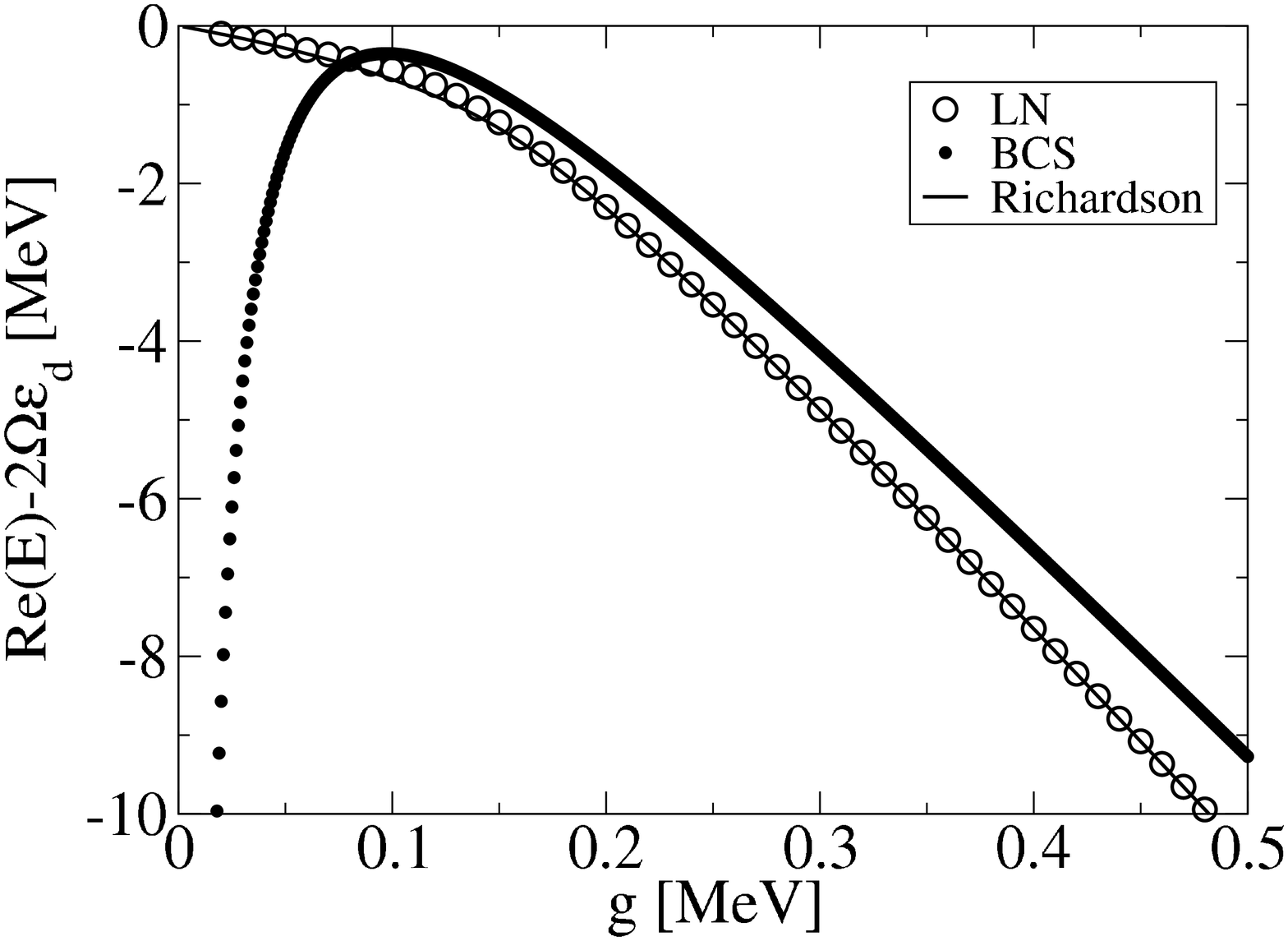}
    \\
    \vspace{10mm}
     \includegraphics[angle=-90,width=0.45\textwidth]{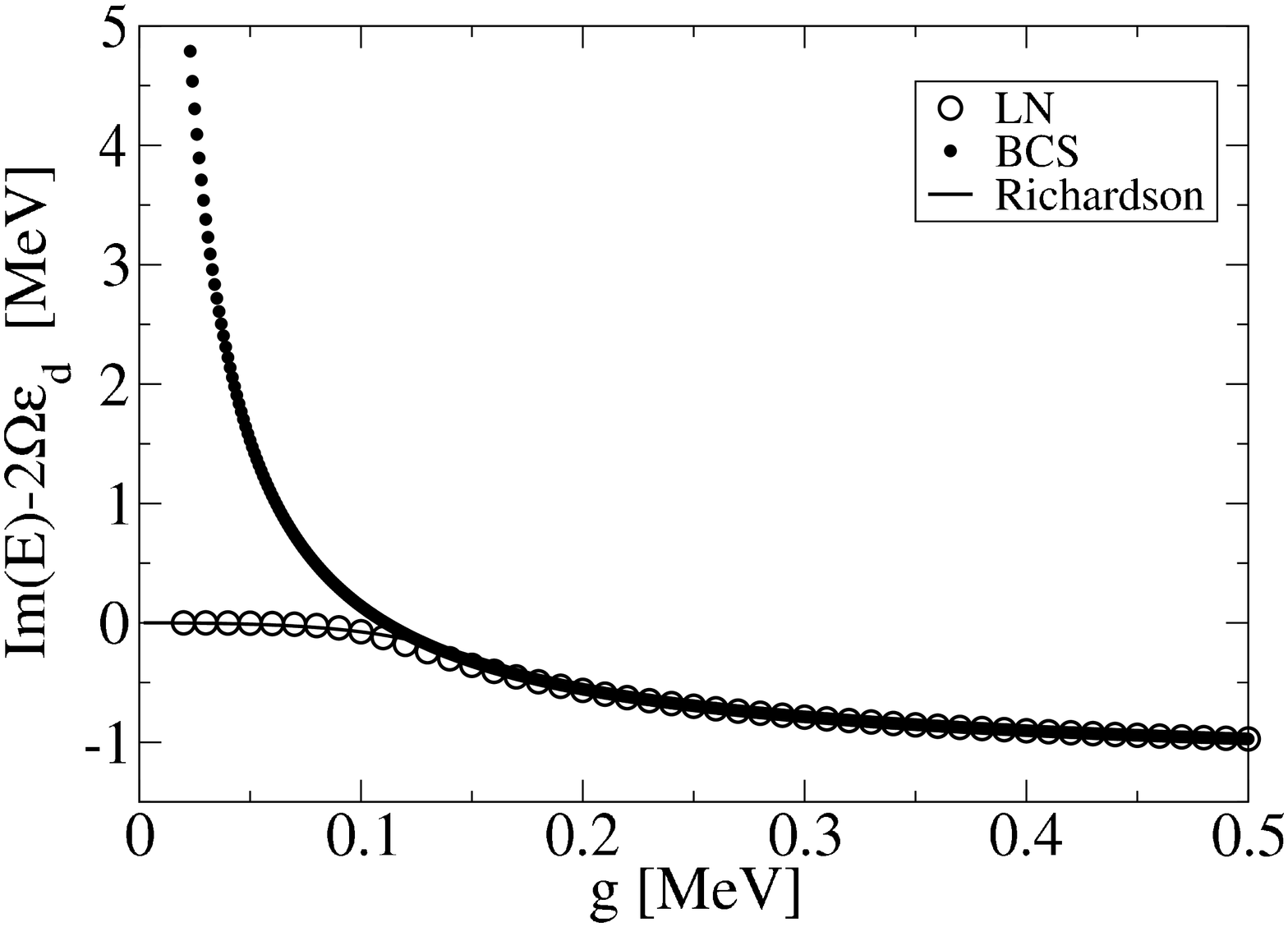}
    \caption{\label{fig.eri} Real (upper graph) and imaginary (lower graph) parts of the ground state energy for $\gamma=0.25$ MeV as a function of the pairing strength $g$ for the BCS, Lipkin-Nogami and Exact solutions.}
\vspace{5mm}
\end{figure}

The ground state energy is calculated in the BCS and LN approximations for $\gamma=0.25$ MeV and shown in Fig. \ref{fig.eri}. The energy of the BCS approximation diverges for values of the strength for which its gap is purely imaginary. While the imaginary part of the energy are similar in both approximations, the real one differs for the same value of the strength. Figure \ref{fig.eri} also compare the real and imaginary parts of the energy with the exact Richardson solution. A good agreement with the LN approximation for all value of $g$ can be observed.

%%%%%%%%%%%%%%%%%%%%%%%%%%%%%%%%%%%%%%%%%%%%
%\section{CONCLUSIONES}\label{sec.conclusiones}
\section{DISCUSSION AND CONCLUSIONS}\label{sec.conclusions}
The Lipkin model with complex energy had been solved in the BCS and LN approximations and compared with the exact Richardson solution. 

The extension of the pairing solution to the complex energy plane, shows that there is solution for any value of the strength, even in the BCS approximation, but this solution is completely nonphysical since the energy diverges.

It was found the LN approximation agreed very well with the exact Richardson solution for real and complex energy for any value of the strength. 

A limitation of the exact Richardson solution is that it can be applied only to constant pairing interaction. But the good agreement with the LN solution seems to indicate that, for more general interactions, the LN method would be a well founded alternative, even in a complex energy representation.

%%%%%%%%%%%%%%%%%%%%%%%%%
%\section{AGRADECIMIENTOS}
\section{ACKNOWLEDGMENTS}
This work has been founded by the National Council of Research (Consejo Nacional de Investigaciones Cient\'ificas y T\'ecnicas - CONICET) by the grant PIP-625.

%-------------
% Bibliography
%\bibliography{richardson}
% \begin{thebibliography}{90}
% \bibitem{1964Cox} J. R. Cox, Journal of Mathematical Physics {\bf 5}, 1065 (1964).
% \end{thebibliography}

\end{document}